\documentclass{article}
\usepackage{arxiv}

\usepackage[utf8]{inputenc} % allow utf-8 input
\usepackage[T1]{fontenc}    % use 8-bit T1 fonts
\usepackage{hyperref}       % hyperlinks
\usepackage{url}            % simple URL typesetting
\usepackage{booktabs}       % professional-quality tables
\usepackage{amsfonts}       % blackboard math symbols
\usepackage{nicefrac}       % compact symbols for 1/2, etc.
\usepackage{microtype}      % microtypography 
\usepackage{lipsum}
\usepackage{amsmath,mathtools,bbm,multirow,url,hyperref,enumerate}
\usepackage{color}
\usepackage{xcolor}
\usepackage{float}
\usepackage{hyperref}
\usepackage{makecell}
\usepackage{algorithm,algorithmic}
\usepackage{multicol}
\usepackage{natbib}
\usepackage{hyperref}
\usepackage{soul}
\hypersetup{colorlinks,citecolor=blue}
\title{Visualizing hypothesis tests in survival analysis under anticipated delayed effects}

\author{
Jos\'e L. Jim\'enez \\
Novartis Pharma A.G. \\
Basel, Switzerland \\
   \And
   Isobel Barrott \\
Janssen \\
UK \\
   \And
Francesca Gasperoni \\
Novartis Pharma A.G. \\
Basel, Switzerland \\
   \And
Dominic Magirr \\
Novartis Pharma A.G. \\
Basel, Switzerland \\
}

\begin{document}
\maketitle
\begin{abstract}
What can be considered an appropriate statistical method for the primary analysis of a randomized clinical trial (RCT) with a time-to-event endpoint when we anticipate non-proportional hazards owing to a delayed effect? This question has been the subject of much recent debate. The standard approach is a log-rank test and/or a Cox proportional hazards model. Alternative methods have been explored in the statistical literature, such as weighted log-rank tests and tests based on the Restricted Mean Survival Time (RMST). While weighted log-rank tests can  achieve high power compared to the standard log-rank test, some choices of weights may lead to type-I error inflation under particular conditions. In addition, they are not linked to an unambiguous estimand. Arguably, therefore, they are difficult to intepret. Test statistics based on the RMST, on the other hand, allow one to investigate the average difference between two survival curves up to a pre-specified time point $\tau$ -- an unambiguous estimand. However, by emphasizing differences prior to $\tau$, such test statistics may not fully capture the benefit of a new treatment in terms of long-term survival. In this article, we introduce a graphical approach for direct comparison of weighted log-rank tests and tests based on the RMST. This new perspective allows a more informed choice of the analysis method, going beyond  power and type I error comparison.

\end{abstract}

% keywords can be removed

\section{Introduction}

Delayed effects are a well-known form of non-proportional hazards typically linked to checkpoint inhibitors in immuno-oncology (IO). In a randomized clinical trial (RCT) comparing chemotherapy with an IO compound, we may well expect the Kaplan-Meier curves to remain close to equal for some time before they diverge. One potential explanation for this behavior is that an IO agent does not target the tumor directly; it instead boosts the patient's immune system, and this positive effect may not be observed immediately. The lag between the activation of immune cells, their proliferation and impact on the tumor is described in the literature as a delayed treatment effect. 

There is an ongoing debate about what should be considered an appropriate  primary analysis for such trials, in particular whether to keep the log-rank test (or the Cox model) as a de-facto standard, or to employ alternative methods   \citep{freidlin2019methods,uno2020log, huang2020estimating,freidlin2020reply}.  Recent publications (e.g., \cite{lin2020alternative,royston2020simulation,chen2020comparison,jimenez2019properties,jimenez2022quantifying, roychoudhury2021robust, mukhopadhyay2022log}) suggest that alternative testing methods, for example weighted log-rank tests, are better tailored for settings with non-proportional hazards, based largely on the fact that they can achieve higher power. This view has been challenged, however, on at least two grounds. Firstly, some of the proposed methods have been shown to produce counter-intuitive results.  For example, \cite{freidlin2019methods} construct a scenario in which the experimental treatment is uniformly worse than control, and yet a particular choice of weighted log-rank test would have a high chance of rejecting the null hypothesis in favour of the experimental treatment. Secondly, weighted log-rank tests are not constructed around an unambiguous estimand. \cite{huang2020estimating} argue that, as a matter of principle, the primary analysis of an RCT should be an estimation procedure with a corresponding confidence interval. Application of this principle would rule out a weighted log-rank test. In the case of non-proportional hazards, \cite{huang2020estimating} recommend to estimate the difference in restricted mean survival time (RMST) between the two arms. Another potential estimand would be the difference in survival probabilities at a specific time, a so-called milestone analysis.   

We do not hope to settle this debate in this article. Instead, our aim is to develop graphical comparisons of the various testing methods, deepening our understanding of how they work.  To do so, we follow \cite{magirr2019modestly} and represent a weighted log-rank test statistic as a difference in mean score between the two treatment arms. We then show that we can express tests that are based on the Kaplan-Meier estimates of the restricted mean survival time (RMST) and milestone survival probabilities in the same way via the concept of ``pseudo-values'' or ``pseudo-observations'' proposed by \cite{andersen2003generalised}. This allows us to put both the weighted log-rank tests and the methods derived from the Kaplan-Meier curves in a common framework which makes them directly comparable. Furthermore, we provide an R package, publicly available on CRAN, to make the proposed methods effectively reproducible.

The rest of the article is organized as follows.  In Section \ref{sc_poplar} we introduce the POPLAR trial that we use to illustrate the common visualization framework proposed in this article. In Section \ref{sc_method}, we introduce the basic theory behind weighted log-rank tests and the tests based on the RMST and milestone survival probabilities, as well as how to represent them as between-arm differences in mean scores and pseudo-values, respectively. In Section \ref{sc_graphical_comparison}, we present the graphical comparison of the various survival tests. We conclude the article in Section \ref{sc_discussion} with a discussion and some final remarks.

\section{The POPLAR trial (NCT01903993)}
\label{sc_poplar}

\begin{figure}[h]
  \centering
   \caption{Kaplan-Meier curves from the POPLAR trial.}
  \vspace{0.5cm}
  \includegraphics[scale=0.6]{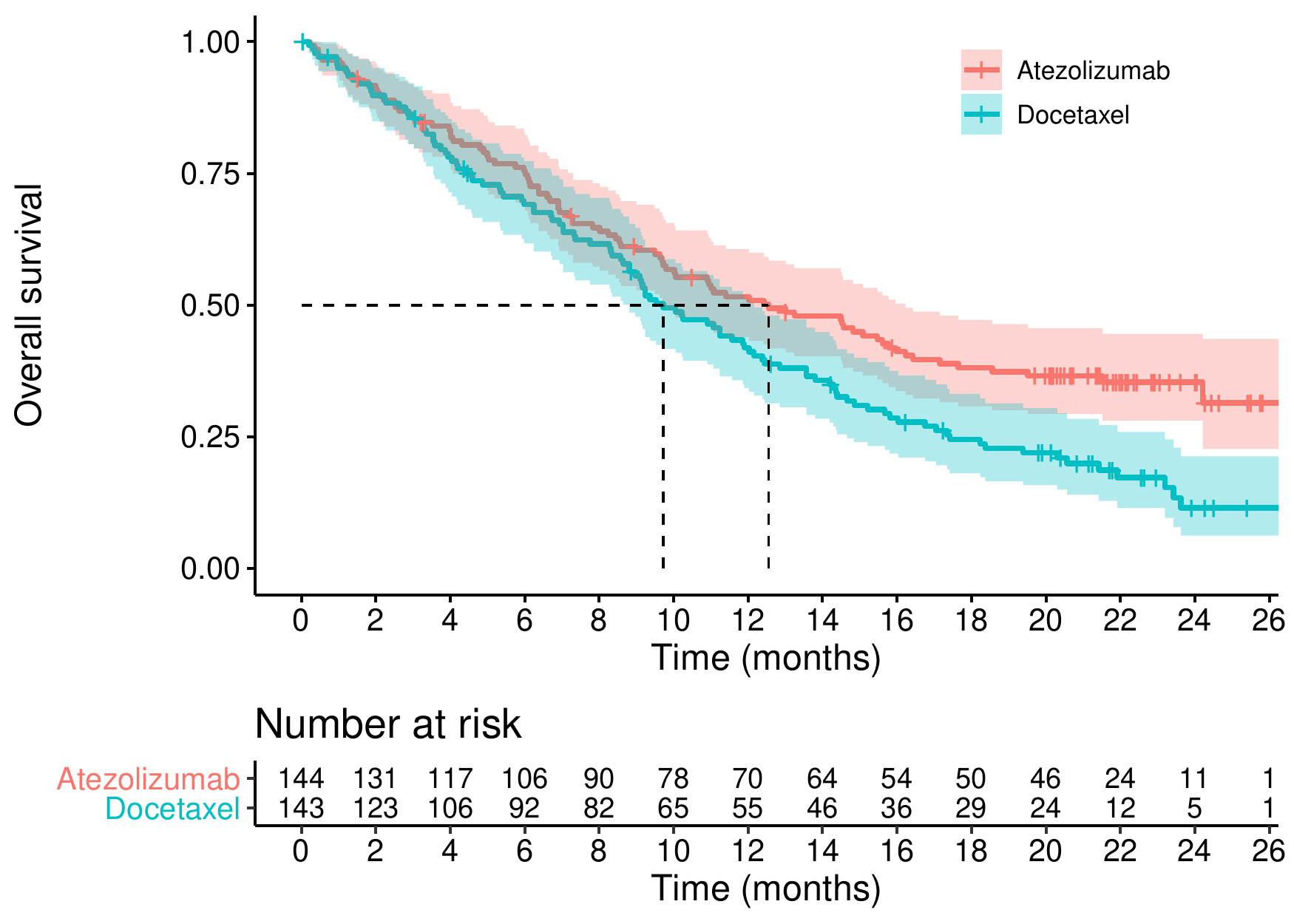}
  \label{figure_kaplan_meier_POPLAR}
\end{figure}

We shall use the POPLAR trial (\cite{fehrenbacher2016atezolizumab}) as a starting point for our discussion. This was an open-label phase 2 randomized controlled trial of atezolizumab versus docetaxel for patients with previously-treated non-small-cell lung cancer. Key design assumptions and de-identified data   are publicly available (\cite{gandara2018blood}). The sample size was calculated assuming a median overall survival (OS) of 8 months for the control arm and a HR of 0.65, which translated into an assumed median OS of approximately 12.3 months for the atezolizumab arm, under an exponential model. Recruitment lasted 8 months. Three interim analyses were planned, with (two-sided) alpha levels of  0.0001, 0.0001, and 0.001. The final analysis of OS was performed when 173 deaths had occurred in the intention-to-treat (ITT) population, using a two-sided $\alpha$ level of 4.88\%. The trial enrolled a total of 287 patients. A Kaplan-Meier estimate derived from the published data set \cite{gandara2018blood} is shown in Figure \ref{figure_kaplan_meier_POPLAR},  where a delayed effect is identifiable.

\section{Methods}
\label{sc_method}

\subsection{Weighted log-rank tests}
\label{sc_weighted_tests}

To implement a weighted log-rank test, we scan over the ordered unique event times $t_1,t_2,\ldots,t_J$, and take a weighted sum of the observed minus expected events on one of the treatment arms, where the expectation is taken assuming that the survival distributions on the two arms are identical. Let $n_{i,j}$ denote the number of patients at risk on treatment $i=0,1$ just prior to time $t_j$, and let $O_{i,j}$ denote the observed number of events on treatment $i=0,1$ at time $t_j$, with the expected number of events given by $E_{i,j} = O_{j}\times n_{i,j} / n_{j}$, where $n_j = n_{0,j}+n_{1,j}$ and $O_j = O_{0,j}+O_{1,j}$. The weighted log-rank test statistic is defined as
\begin{equation*}
    U_W := \sum_{j=1}^J w_j\left( O_{1,j} - E_{1,j}\right) \sim  N(0, V_W),
\end{equation*}
where
\begin{equation*}
    V_W = \sum_{j=1}^J w_j^2\frac{n_{0,j}n_{1,j}O_j(n_j - O_j)}{n_j^2(n_j - 1)},
\end{equation*}and $w_j$ represents the weight at time $t_j$.

If the experimental treatment is beneficial compared to the control treatment, we should see fewer events on the experimental arm than would be expected assuming identical survival curves on the two treatments, leading to $U_W << 0$. The one-sided p-value is calculated as $p_W =\Phi(U_W / \sqrt{V_W})$, where $\Phi(.)$ represents the cumulative distribution function of the standard Normal distribution.

A weighted log-rank test allows one to pre-specify the weights to boost the chances that $p<\alpha$, given the expected treatment effect. In other words, because of the anticipated delayed Kaplan-Meier curves separation, we will choose the weights in a way that early events, that occur when the Kaplan-Meier curves are expected to be similar to each other, are underweighted compared to late events. \cite{fleming2011counting} proposed a class of weight functions with $w_j = ( \hat{S}(t_j-))^{\rho} \times ( 1 - \hat{S}(t_j-))^{\gamma} $, where $\hat{S}(t_j-)$ is the Kaplan-Meier estimate of the pooled sample just prior to time $t_j$. Under anticipated delayed treatment effects, a popular choice is the Fleming-Harrington-(0,1) test, which uses $\rho=0$, $\gamma = 1$ and therefore the weight $w_j = 1 - \hat{S}(t_j-)$. A recently proposed alternative to the Fleming-Harrington class of weights is the modestly-weighted log-rank test (see \cite{magirr2019modestly}), which uses $w_j = 1 / \max{\left\lbrace \widehat{S}(t_j-), s^*\right\rbrace}$, for fixed $s^*$. In common with the Fleming-Harrington-(0,1) test, it gives relatively larger weight to later events than to earlier events, but it has the advantage that it controls the type-I error rate not only for the sharp null hypothesis $H_0: S_1(t)=S_0(t) ~\forall ~t$ but also for the strong null $\tilde{H}_0: S_1(t) \leq S_0(t)~\forall ~t$. The first event receives a weight of 1, and weights are increasing up to a maximum $1/s^*$. So choosing $s^*=0.5$, for example, allows the weights to range from 1 to 2 (assuming that event rates are such that $\widehat{S}(t) = 0.5$ for some $t$) which may be a reasonable choice in many immuno-oncology trials. Choosing $s^*\approx 1$  makes the test more similar to a standard log-rank test.  A tutorial on how to design a randomized clinical trial using the modestly weighted log-rank test was recently published by \cite{magirr2022design}.

Although weighted log-rank tests are usually thought of as a weighted sum of observed minus expected events, they can also be expressed as a difference in mean score on the two treatment arms \cite{leton2001equivalence, magirr2021non}. Letting $l_{1,j}$ and $l_{0,j}$ denote the number of patients censored on the test treatment and control treatment, respectively, during $\left[\left.t_j, t_{j+1}\right)\right.$, we can express the weighted log-rank statistic as

\begin{equation}\label{eqn_w_s}
\begin{split}
U_W &:=\sum_{j = 1}^{J} w_j\left( O_{1,j} - O_j\frac{n_{1,j}}{n_j} \right)\\
  &=\sum_{j = 1}^{J} w_jO_{1,j} - \sum_{j = 1}^{J}w_j\frac{O_{j}}{n_j} \times n_{1,j}\\
  &=\sum_{j = 1}^{J} w_jO_{1,j} - \sum_{j = 1}^{J}w_j\frac{O_{j}}{n_j} \times \sum_{i = j}^{J}(O_{1,i} + l_{1,i})\\
  &=\sum_{j = 1}^{J} w_jO_{1,j} - \sum_{j = 1}^{J}(O_{1,j} + l_{1,j}) \times \sum_{i = 1}^{J}w_i\frac{O_{i}}{n_i}\\
  &=\sum_{j = 1}^{J} O_{1,j}\left( w_j - \sum_{i = 1}^{j}w_i\frac{O_{i}}{n_i} \right) +  \sum_{j = 1}^{J}l_{1,j} \left( - \sum_{i = 1}^{j}w_i\frac{O_{i}}{n_i}\right).\\
\end{split}    
\end{equation}

The final expression in (\ref{eqn_w_s}) can also be written as a sum of "scores", $a_k$, over all patients $k=1,\ldots,N$
\begin{equation*}
    U_W = \sum_{k= 1}^N \mathbb{I}\left\lbrace z_{k} = 1 \right\rbrace a_{k}
\end{equation*}
where $z_k$ denotes the treatment assignment of patient $k$.  If patient $k$ has an event at time $t_j$, then they are given a score of $a_k:=w_j - \sum_{i = 1}^{j}w_i\frac{O_{i}}{n_i}$, where the summation is over the observed event times up to time $t_j$. If, instead, patient $k$ had a censored observation during $\left[\left.t_j, t_{j+1}\right)\right.$ they would receive a score of $a_k:=-\sum_{i = 1}^{j}w_i\frac{O_{i}}{n_i}$. 

The estimator $V_W$ for the variance of $U_W$ is usually derived from a sampling viewpoint. It can be arrived at via a Cox model with treatment term only. When we fit a Cox model, we consider the variance of $U_W$ under repeated iterations of the experiment from the same (super)population. When we express $U_W$ as a sum of scores, however, it shifts us towards a randomization inference perspective. That is, we consider the sample fixed, and the only source of randomness is the treatment assignment. Under this framework, and indexing the patients such that the first $k=1,\ldots,N_1$ patients are those on treatment $1$, another way to express $U_W$ is 
\begin{equation*}
U_W = \sum_{k = 1}^{N_1} A_l
\end{equation*}
where  the $A_1,\ldots,A_{N_1}$ are a random sample of size $N_1$ from $a_1,\ldots,a_N$. In particular, 
\begin{equation*}
    p(A_k = a) = \frac{1}{N},\text{ for }a= a_1,\ldots,a_N
\end{equation*}
and
\begin{equation*}
    p(A_k = a, A_l = a') = \left\lbrace \begin{array}{c c} \frac{1}{N(N-1)} & \text{ for }a\neq a' \\ 0 & \text{otherwise.} \end{array}  \right.
\end{equation*}
Also note that $\sum_{k= 1}^N a_k = 0$, as can be seen by switching the treatment labels in (\ref{eqn_w_s}), 
\begin{equation}
    \begin{split}
     \sum_{k= 1}^N \mathbb{I}\left\lbrace z_{k} = 1 \right\rbrace a_{k} +  \sum_{k= 1}^N \mathbb{I}\left\lbrace z_{k} = 0 \right\rbrace a_{k} &= \sum_{j = 1}^{J} w_j\left( O_{1,j} - O_j\frac{n_{1,j}}{n_j} \right)+ \sum_{j = 1}^{J} w_j\left( O_{0,j} - O_j\frac{n_{0,j}}{n_j}\right) \\
     &= 0.
     \end{split}
\end{equation}
Thus,
\begin{equation*}
    E(A_k) = \frac{1}{N}\sum_{i=1}^N a_k = 0,
\end{equation*}
\begin{equation*}
    Var(A_k) = \frac{1}{N}\sum_{i=1}^N a_k^2,
\end{equation*}
and
\begin{equation*}
  \begin{split}
    Cov(A_k,A_l) &= E(A_k A_l) \\
                 &= \frac{1}{N(N-1)}\sum_{k\neq l} a_ka_l \\
                 &= \frac{1}{N(N-1)}\left\lbrace \left(\sum_{k=1}^N a_k\right)^2 - \sum_{k=1}^N a_k^2 \right\rbrace \\
                 &= \frac{-1}{N(N-1)}\sum_{i=1}^N a_k^2.
  \end{split}
\end{equation*}
It follows that $E_P(U_W) = 0$ and
\begin{equation*}
  \begin{split}
    Var_P(U_W) &= \sum_{k = 1}^{N_1} Var(A_k) + \sum_{k \neq l} Cov(A_k, A_l) \\
    &= \frac{N_1}{N}\sum_{k=1}^N a_k^2 - \frac{N_1^2 - N_1}{N(N-1)}\sum_{k=1}^N a_k^2\\
    &= \frac{N_1(N-1) - (N_1^2 - N_1) }{N(N-1)}\sum_{k=1}^N a_k^2 \\
    &= \frac{N_1(N - N_1) }{N(N-1)}\sum_{k=1}^N a_k^2,
  \end{split}
\end{equation*}
where we use the subscript $P$ to emphasize this refers to the permutation distribution of $U_W$.
Note that instead of using $U_W$, we could just as well use the difference in mean score on the two arms,
\begin{equation}\label{eqn_u_tilde}
  \begin{split}
    \tilde{U}_W &= \frac{\sum_{k= 1}^N \mathbb{I}\left\lbrace z_{k} = 1 \right\rbrace a_{k}}{N_1} - \frac{\sum_{i= 1}^N \mathbb{I}\left\lbrace z_{k} = 0 \right\rbrace a_{k}}{N- N_1}\\
    &= \frac{U_W}{N_1} - \frac{\sum_{k=1}^N a_k - U_W}{N-N_1}\\
    &= U_W\left( \frac{1}{N_1} + \frac{1}{N-N_1}\right)
    \end{split}
\end{equation}
as the test statistic, with $E_P(\tilde{U}_W) = 0$ and $Var_P(\tilde{U}_W)  = (N_1^{-1}+(N-N_1)^{-1})^2Var_P(U_W)$.

It can be shown that, asymptotically, $\tilde{U}_W\sim N(0, Var_P(\tilde{U}_W))$, and this could be used for inference. Alternatively, one could compute an exact p-value by considering all permutations of the treatment labels. Also, although $V_W$ and $Var_P(U_W)$ are derived via different frameworks, it can be shown that they are approximately the same when the censoring distribution is the same on the two arms \citep{leton2001equivalence}, as should usually be the case for administrative censoring in an RCT. One may perform the weighted log-rank test with variance estimator $V_W$ but keeping in mind the permutation interpretation to gain insight into how the test statistic behaves.

To make it easier to visually compare scores from different weighted tests we can also perform a re-scaling
\begin{equation*}
b_k = \frac{2a_k - \max{a} - \min{a}}{\max{a} - \min{a}}
\end{equation*}
so that the standardized scores $b_k \in (-1,1)$. This leaves the p-value of the permutation test unchanged.

\subsection{Tests based on the Restricted Mean Survival Time (RMST) and Milestone survival rates}
\label{sc_pseudo_values}

The $\tau$-restricted mean survival time (RMST) is defined as $\mbox{RMST}(\tau) = E(\min \left\lbrace T,\tau \right\rbrace ) = \int_{0}^{\tau} S(t)dt $, where $S(t) = P(T > t) = \exp \left ( -\int_{0}^{t} \lambda(u) du \right )$ is the survival function and $\lambda(t)$ the hazard function (\cite{royston2011use,tian2014predicting}). The RMST represents the expected lifetime over a time horizon equal to $\tau$, that in the context of the primary analysis of a clinical trial would have to be pre-specified. The test statistic linked to the difference between the RMST of the experimental and control arms is usually based on the Kaplan-Meier estimator and calculated as follows. With notation introduced in Section \ref{sc_weighted_tests}, let $U_{\tiny \mbox{RMST}} = \int_{0}^{\tau} \left ( \widehat{S}_{\tiny \mbox{1}}(t) - \widehat{S}_{\tiny \mbox{0}}(t) \right ) dt$ be the difference in estimated RMST on treatment and control arms, respectively, and $V_{\tiny \mbox{RMST}} = \sum_{i=0,1} \sum_{j=1}^{J} (\int_{t_j}^{\tau} \widehat{S}_i(t)dt)^2 \frac{O_{i,j}}{n_{i,j}(n_{i,j}-O_{i,j})}$ its variance. The one-sided p-value is calculated as $p_{\tiny \mbox{RMST}} = \Phi(-U_{\tiny \mbox{RMST}} / \sqrt{V_{\tiny \mbox{RMST}}})$.

Milestone survival analysis (\cite{klein2007analyzing}) compares survival probabilities at a fixed point in time $\kappa$. Let $U_{\tiny \mbox{MLST}} = \widehat{S}_{1}(\kappa) - \widehat{S}_{0}(\kappa)$ be the difference between the  estimated survival probabilities at time $\kappa$ on treatment and control arms, respectively, and $V_{\tiny \mbox{MLST}} = \sum_{i=0,1}\sum_{t_j \leq \kappa} \widehat{S}_i(\kappa)^2 \frac{O_{i,j}}{n_{i,j}(n_{i,j}-O_{i,j})}$ its variance. The one-sided p-value is calculated as $p_{\tiny \mbox{MLST}} = \Phi(-U_{\tiny \mbox{MLST}} / \sqrt{V_{\tiny \mbox{MLST}}})$. 

In these forms, the RMST and the milestone analysis are difficult to compare with a weighted log-rank test. However, these two analyses, that traditionally employ Kaplan-Meier estimators, can also be addressed with the use of pseudo-values (\cite{andersen2003generalised}) in such a way that the pseudo-values are directly analogous with the scores derived for the weighted log-rank test.

For the RMST($\tau$), the $k$-th pseudo-value, $k=1,\ldots,N$, is defined as 
\begin{equation}
\label{eq_pseudo_rmst}
    \theta_{k}^{\tiny \mbox{RMST},\tau} = N \int_0^\tau \widehat{S}(t) dt - (N-1) \int_0^\tau \widehat{S}^{(-k)}(t) dt,
\end{equation}where $\widehat{S}^{(-k)}(t)$ is the Kaplan-Meier estimator excluding patient $k$. It can be shown \citep{graw2009pseudo, jacobsen2016note, overgaard2017asymptotic} that, asymptotically, 
\begin{equation}
    E(\theta_{k}^{\tiny \mbox{RMST}, \tau} \mid Z_k) \approx E(\min\left\lbrace T_k ,\tau\right\rbrace \mid Z_k),
\end{equation}
where $Z_k$ denotes treatment assignment. Thus, one may estimate the difference in RMST($\tau$) between the two arms as
\begin{equation*}
    \tilde{U}_{\tiny \mbox{RMST},\tau} = \frac{\sum_{k= 1}^N \mathbb{I}\left\lbrace z_{k} = 1 \right\rbrace \theta_{k}^{\tiny \mbox{RMST},\tau}}{N_1} - \frac{\sum_{k= 1}^N \mathbb{I}\left\lbrace z_{k} = 0 \right\rbrace \theta_{k}^{\tiny \mbox{RMST},\tau}}{N- N_1}
\end{equation*}
For the purpose of testing the null hypothesis $H_0: S_1(t) = S_0(t) ~\forall~ t$, the statistic $\tilde{U}_{\tiny \mbox{RMST}}$ is directly analagous to $\tilde{U}_W$ in (\ref{eqn_u_tilde}), with the pseudovalues $ \theta_{k}^{\tiny \mbox{RMST},\tau}$ playing the role of the $a_k$.
It is usually recommended to estimate the variance of $\tilde{U}_{\tiny \mbox{RMST}}$ via non-parametric bootstrap or using a robust variance estimator for a linear model. In principle, for the purpose of testing the null hypothesis $H_0$, one could switch to a randomization inference perspective and consider the permutation distribution of $\tilde{U}_{\tiny \mbox{RMST}}$, as has been previously suggested by \cite{horiguchi2020permutation}. In this case, one is free to re-scale the pseudo-values without it changing the p-value. As mentioned in Section \ref{sc_weighted_tests}, one does not have to actually implement the permutation test in order to benefit from the additional insight it provides. By comparing (re-scaled versions) of $\theta_{k}^{\tiny \mbox{RMST},\tau}$ and $a_k$, we can better understand the behaviour of the respective tests. It is important to note, of course, that the usefulness of $\tilde{U}_{\tiny \mbox{RMST},\tau}$ goes beyond testing the null hypothesis, since it provides us with a point estimate and confidence interval for a meaningful estimand (although one would typically use the Kaplan-Meier based estimators instead). We shall pick up on this aspect in the Section \ref{sc_discussion}. 

For the milestone analysis, the procedure is exactly the same as for RMST, but with pseudo values defined as
\begin{equation}
\label{eq_pseudo_mlst}
    \theta_{k}^{\tiny \mbox{MLST},\kappa} = n \widehat{S}(\kappa) - (n-1) \widehat{S}^{(-k)}(\kappa).
\end{equation}

We shall furthermore consider two novel estimands that can be estimated in the same way via the Kaplan-Meier estimator. Firstly, the between-arm difference in window mean survival time (\cite{paukner2021window}), otherwise known as long-term restricted mean survival time (\cite{horiguchi2023assessing}). The  window mean survival time (WMST) is defined as WMST($\tau_1$, $\tau_2$) = RMST($\tau_2$) - RMST($\tau_1$) for $\tau_2 > \tau_1$. It has the interpretation of the mean amount of time spent alive between $\tau_1$ and $\tau_2$. The second novel estimand proposed by \cite{uno2023ratio} and \cite{snapinn2023treatment} is the between-arm difference in the (log) average hazard with survival weight (AHSW). The AHSW is defined as AHSW($\tau$) = (1 - S($\tau$)) / RMST($\tau$). It can be interpreted as "the average person-time incidence rate of $T$ on $t\in(0,\tau)$ when all $T$ before $\tau$ would have been observed without being censored by study-specific censoring time" (\cite{uno2023ratio}). For both of these novel estimands, the procedure for calculating pseudo-values and estimating between-arm differences is analogous to the procedure for RMST. 

The concept of pseudo-values can also be extended to parametric estimators of the survival curve \cite{nygaard2020regression}, simply by replacing the Kaplan-Meier estimator $\widehat{S}(t)$ with its parametric equivalent. This is something we shall also explore in Section \ref{sc_graphical_comparison}.

\section{Graphical comparison}
\label{sc_graphical_comparison}

In this section, we use the POPLAR trial data described in Section \ref{sc_poplar} to run a series of survival tests and observe how they behave under the lens of the (re-scaled) scores for the weighted log-rank tests, and the (re-scaled) pseudo-values for the tests based on the RMST and Milestone survival rates. By comparing tests in a number of ways we derive several insights.

\subsection{Modestly weighted log-rank tests}

\begin{figure}[h]
  \centering
   \caption{Graphical representation of a modestly-weighted log-rank test, as compared to the log-rank test and milestone analysis, when applied to the POPLAR data. A standardized score is plotted for each patient, where censored observations are semi-transparent. Horizontal dashed lines are plotted at the mean score on each arm. The distance between these two lines corresponds to the numerator in the permutation-based test statistic.}
  \vspace{0.5cm}
  \includegraphics[scale=0.7]{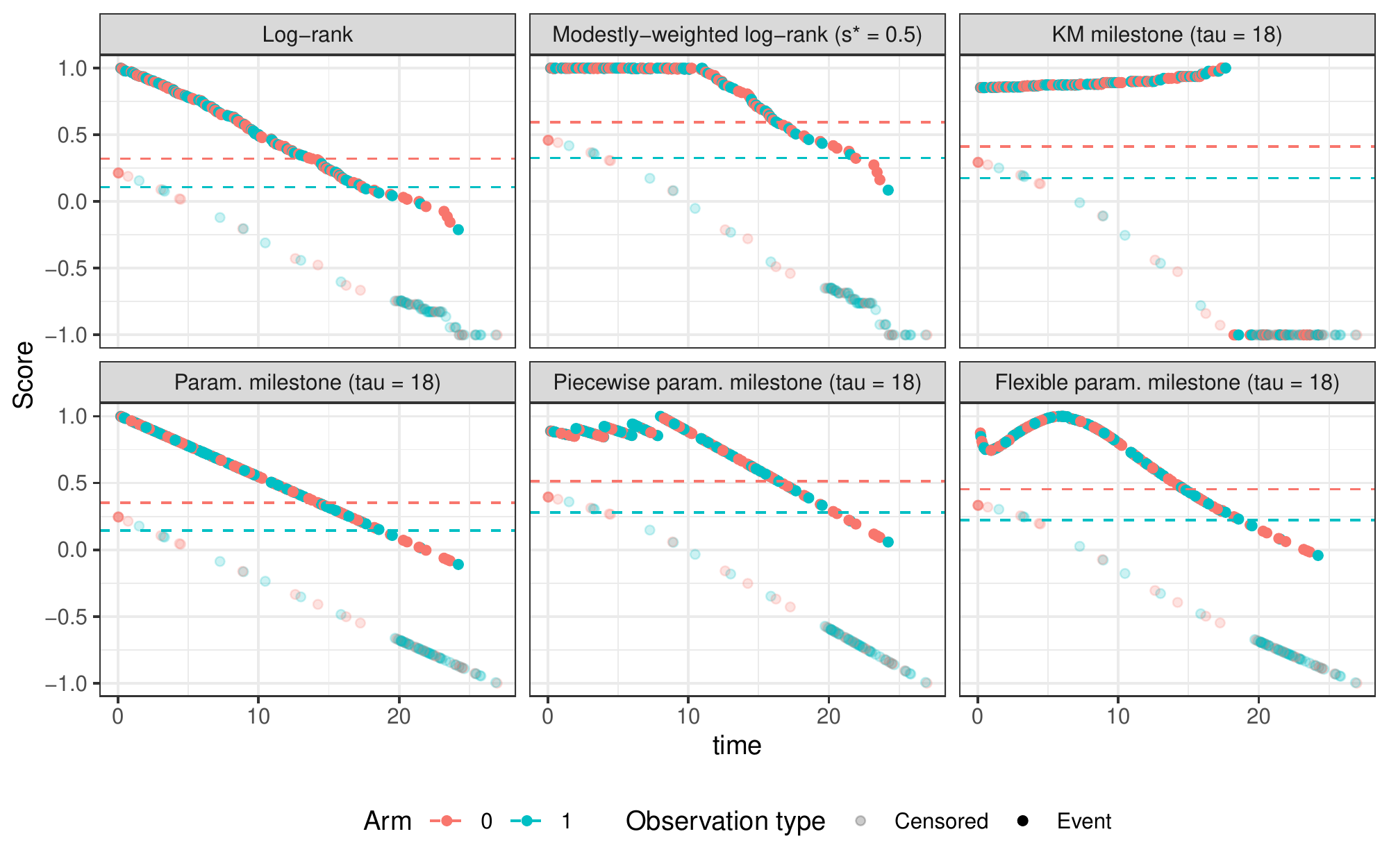}
  \label{fig_modest}
\end{figure}

In Figure \ref{fig_modest}, we plot the scores from the modestly-weighted log-rank test (with $s^* = 0.5$) side-by-side with those from the standard log-rank test, and also with those from Kaplan-Meier and parametric-based milestone analysis at 18 months. The modestly-weighted test appears to be a compromise between the log-rank test and the Kaplan-Meier based milestone analysis, in the sense that the scores for observed events are approximately equal to 1 for an initial period (similar to the milestone analysis) before steadily decreasing for later follow-up times (similar to a log-rank test). With the Kaplan-Meier based milestone analysis there is a sharp dichotomy; events just before and just after the milestone are given extreme opposite scores. The dangers of outcome dichotomization in terms of loss of information are well documented \citep{senn2005dichotomania, fedorov2009consequences}. One way to avoid this loss of information is to apply parametric models instead of the non-parametric Kaplan-Meier estimator \cite{nygaard2020regression}. Three model-based approaches are shown in the bottom row of Figure \ref{fig_modest}. The first uses an exponential model, producing scores that look almost identical to the log-rank scores. The second uses a piecewise exponential model with breakpoints at 2, 4, 6 and 8 months, while the third uses a flexible parametric model with a cubic spline on the baseline cumulative hazard \citep{jackson2016flexsurv}. Both of these more flexible models produce scores that look remarkably like those of the modestly-weighted test.

\subsection{RMST vs log-rank}

\begin{figure}[h]
  \centering
   \caption{Graphical representation of a test based on the difference in restricted mean survival time (RMST) using Kaplan-Meier derived pseudovalues, as compared to the log-rank test and the Fleming-Harrington-(0,1) test, when applied to the POPLAR data. A standardized score is plotted for each patient, where censored observations are semi-transparent. Horizontal dashed lines are plotted at the mean score on each arm. The distance between these two lines corresponds to the numerator in the permutation-based test statistic.}
  \vspace{0.5cm}
  \includegraphics[scale=0.7]{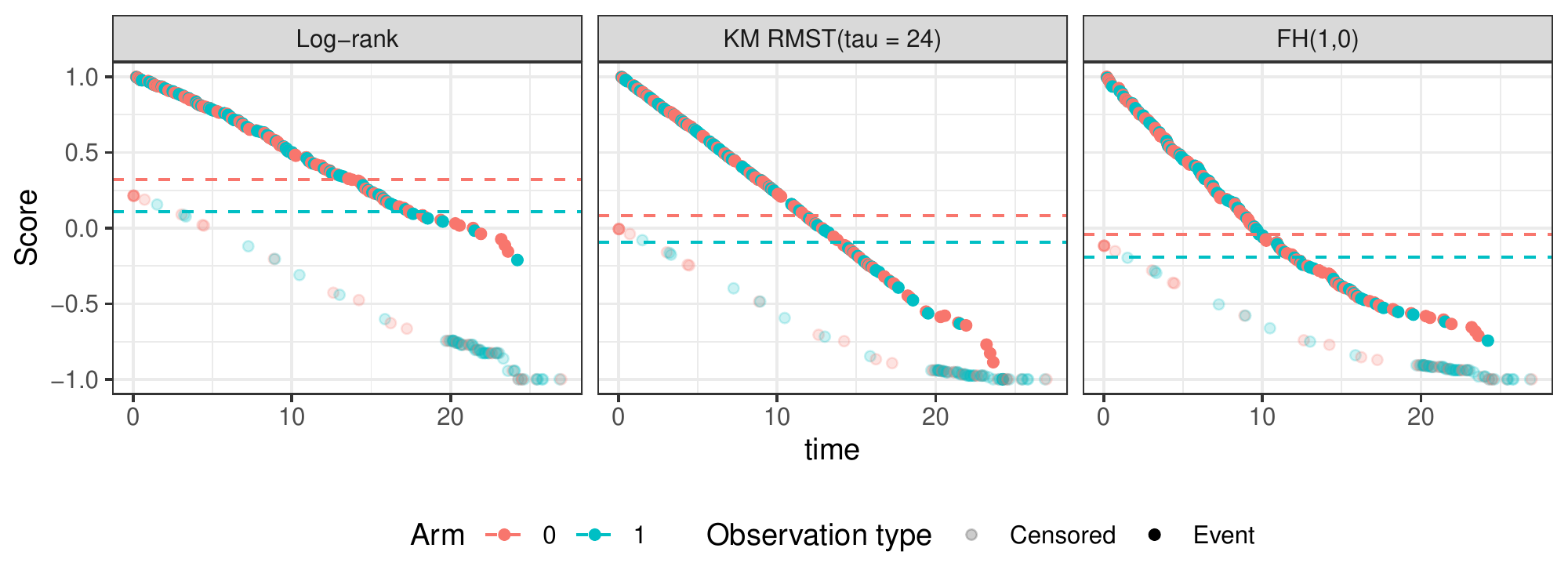}
  \label{fig_rmst}
\end{figure}

In Figure \ref{fig_rmst}, we plot the scores from the Kaplan-Meier based RMST(24) analysis side-by-side with those from the standard log-rank test, and also with those from a Fleming-Harrington-(1,0) test. The RMST test looks more similar to the Fleming-Harrington-(1,0) test than to the standard log-rank test. When interpreting the graphs, the relative vertical distance between a point and the average score lines is  indicative of how influential that point is. For the RMST and FH(1,0) tests these distances are greater at earlier timepoints than at later timepoints, whereas for the log-rank test these distances are greater at later timepoints. So the RMST and FH(1,0) tests are giving greater emphasis to early follow-up times, relative to the log-rank test.

\subsection{Tests based on novel estimands}

\begin{figure}[h]
  \centering
   \caption{Graphical representation of tests based on the difference in window mean survival time (WMST) and (log) average hazard with survival weight (AHSW) using Kaplan-Meier derived pseudovalues,  when applied to the POPLAR data. A standardized score is plotted for each patient, where censored observations are semi-transparent. Horizontal dashed lines are plotted at the mean score on each arm. The distance between these two lines corresponds to the numerator in the permutation-based test statistic.}
  \vspace{0.5cm}
  \includegraphics[scale=0.7]{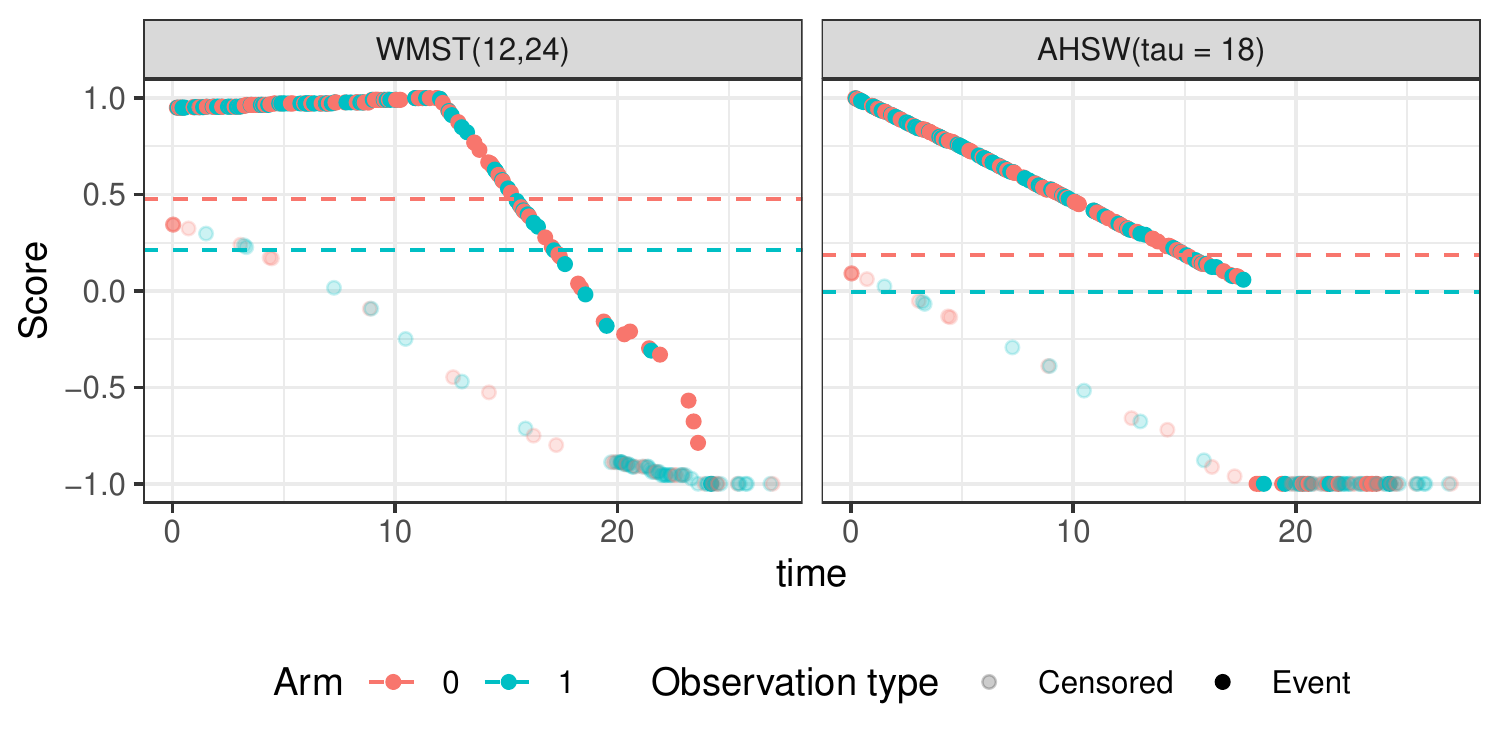}
  \label{fig_novel}
\end{figure}

In Section \ref{sc_pseudo_values}, we described two novel estimands that could be estimated via pseudo-values derived from the Kaplan-Meier estimator of survival in the full data. These were the difference in window-mean survival time (WMST) and the difference in average hazard with survival weight (AHSW). Standardized pseudo-values that could be used to test the null hypothesis $H_0$ are plotted in Figure \ref{fig_novel}. The WMST-based test looks somewhat similar to the modestly-weighted test, except that late censored observations are given the same score as late events. We see the same thing when comparing the AHSW-based test with the log-rank test.

\subsection{Fleming-Harrington-($\rho$, $\gamma$) tests with $\gamma > 0$}

\begin{figure}[h]
  \centering
   \caption{Graphical representation of Fleming-Harrington-($\rho$,$\gamma$) tests, when applied to the POPLAR data. A standardized score is plotted for each patient, where censored observations are semi-transparent. Horizontal dashed lines are plotted at the mean score on each arm. The distance between these two lines corresponds to the numerator in the permutation-based test statistic.}
  \vspace{0.5cm}
  \includegraphics[scale=0.7]{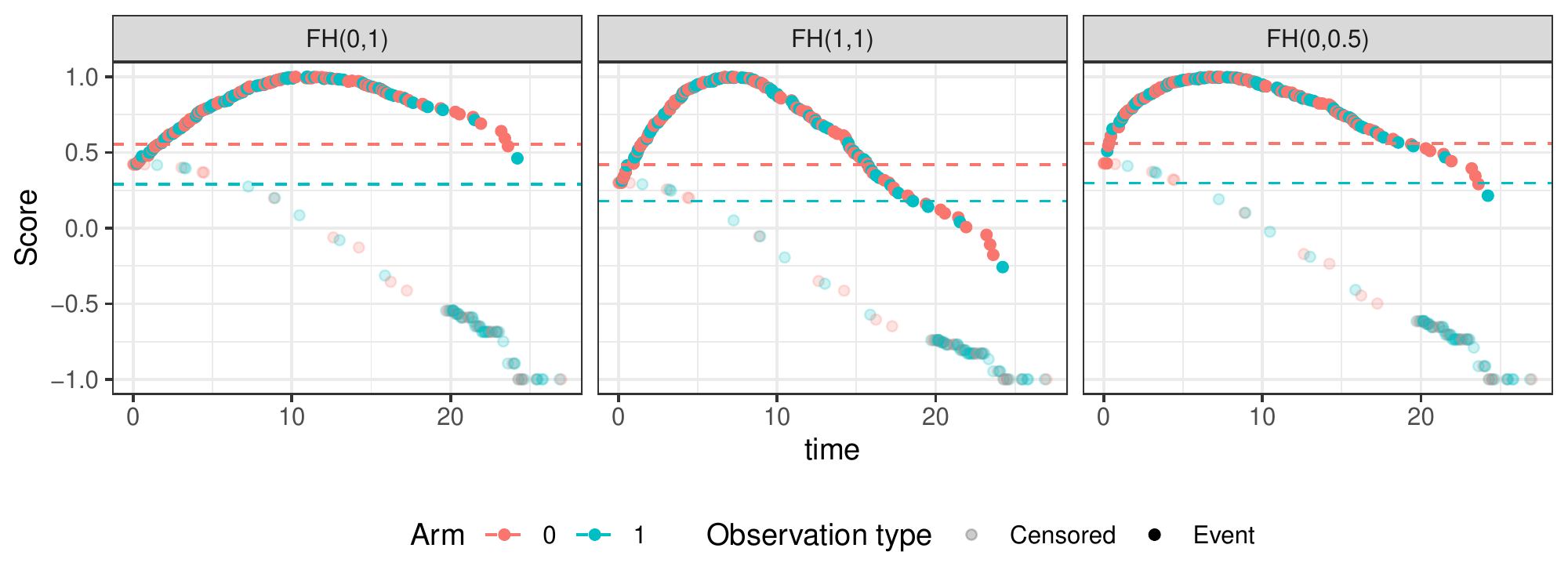}
  \label{fig_fh}
\end{figure}

In Figure \ref{fig_fh}, we plot the scores from a series of Fleming-Harrington-($\rho$, $\gamma$) tests with $\gamma >0$. The pattern of the scores is strikingly different to the other tests in that they are highly non-monotonic in time. The scores for the group of patients with early events are lower, i.e., better, than the scores of many patients with later events. This leads to the strange behaviour and lack of strong type 1 error control that has previously been observed when using these test statistics \cite{freidlin2019methods, magirr2021non}. 

\subsection{Effect of early censoring}

\begin{figure}[h]
  \centering
   \caption{Kaplan-Meier plot from the POPLAR trial data with artificial independent censoring.}
  \vspace{0.5cm}
  \includegraphics[scale=0.7]{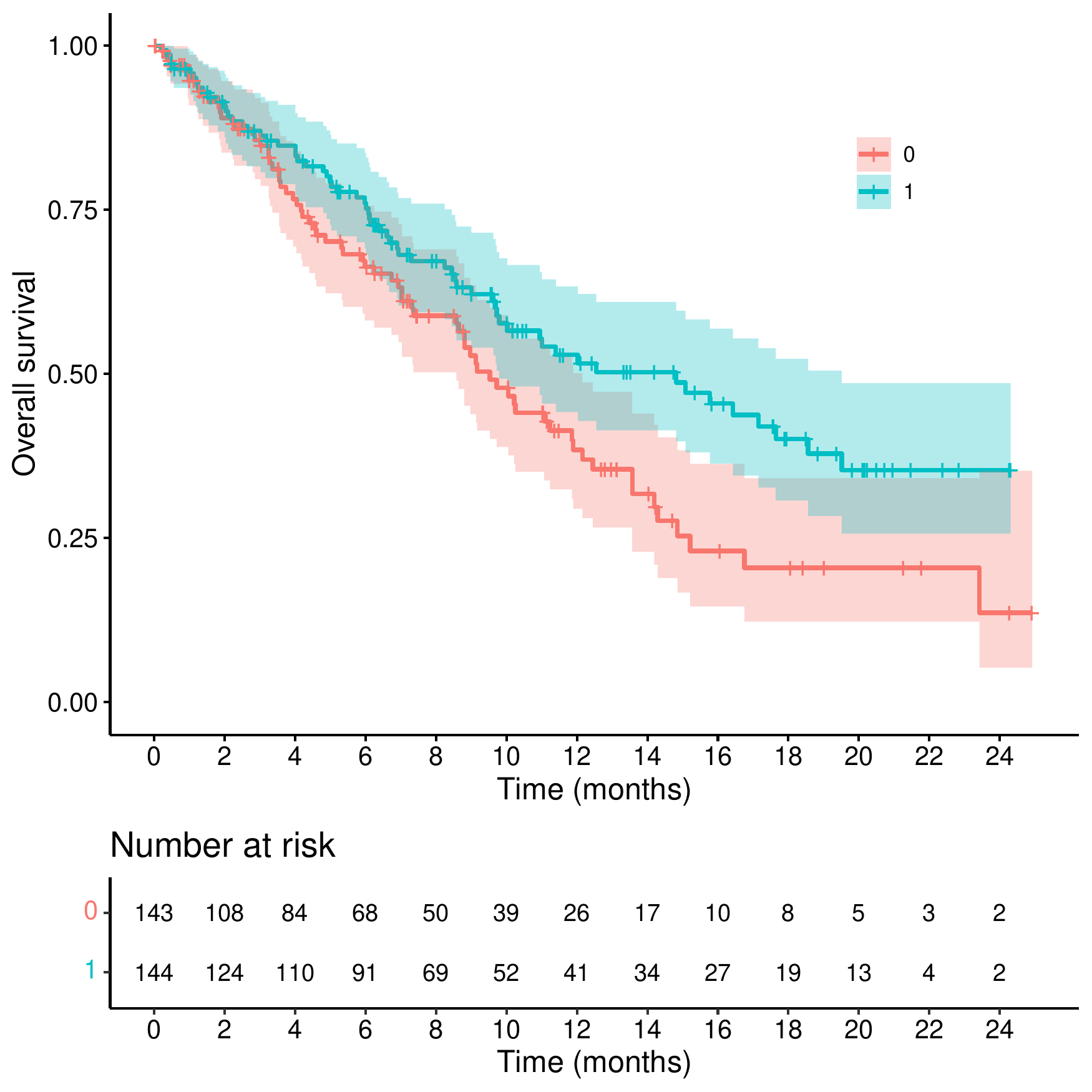}
  \label{fig_km2}
\end{figure}

\begin{figure}[h]
  \centering
   \caption{Graphical representation of selected test statistics, when applied to the POPLAR data with additional artificial censoring, as depicted in Figure \ref{fig_km2}. A standardized score is plotted for each patient, where censored observations are semi-transparent. Horizontal dashed lines are plotted at the mean score on each arm. The distance between these two lines corresponds to the numerator in the permutation-based test statistic.}
  \vspace{0.5cm}
  \includegraphics[scale=0.7]{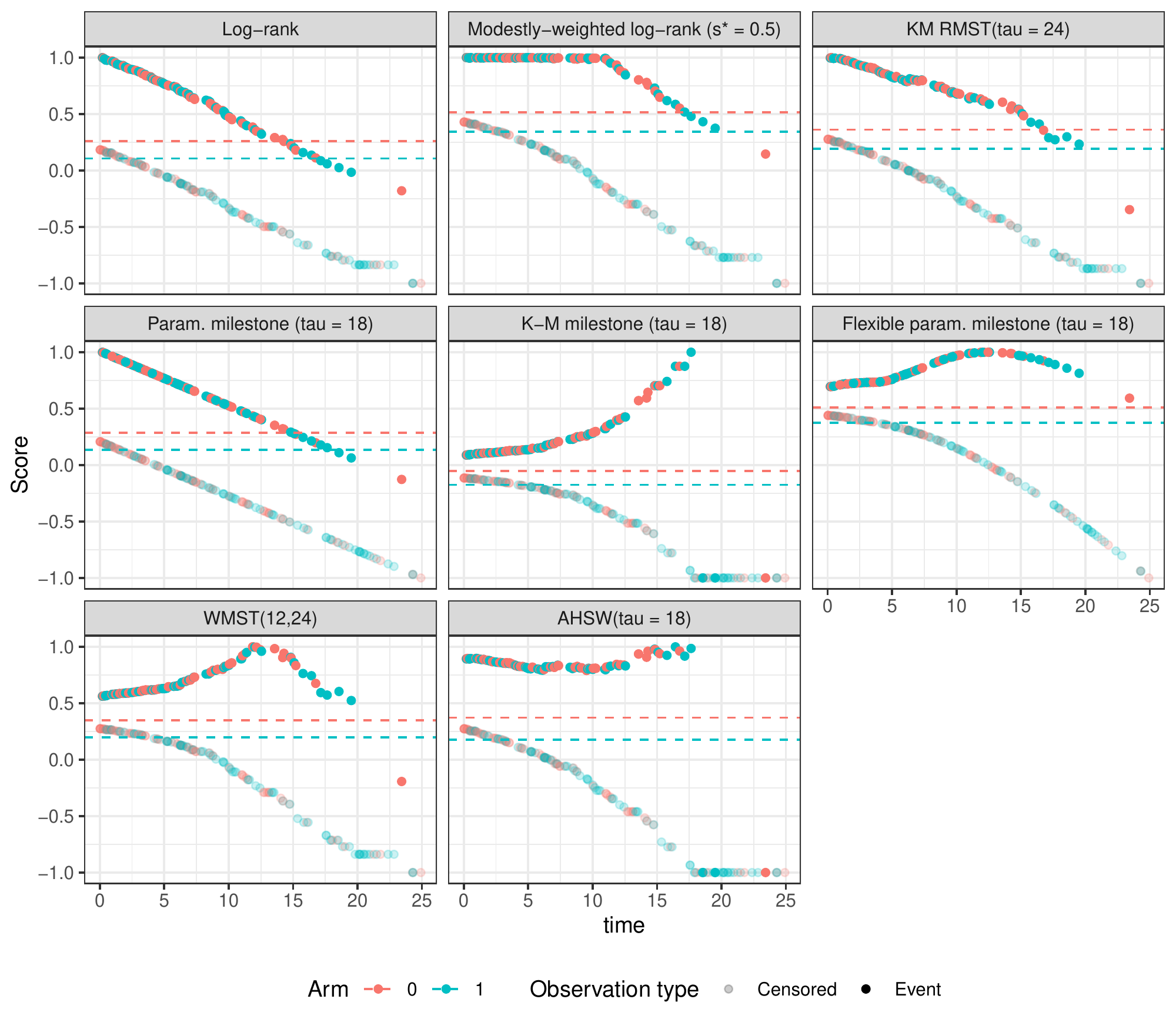}
  \label{fig_cens}
\end{figure}

It is clear from Figure \ref{figure_kaplan_meier_POPLAR}, and from the score plots, that the POPLAR trial had little early censoring. The vast majority of censoring events occur after 20 months, suggesting recruitment lasted approximately 6 months. It may be interesting, however, to understand the behaviour of the test statistics when there is more heavy early censoring. For example, had the recruitment rate been very slow. For this purpose, we artificially mimick a 26 month recruitment period by simulating a competing censoring event from a uniform distribution on (0, 26) months. The resulting data set is shown in Figure \ref{fig_km2}. Then, in Figure \ref{fig_cens}, we repeat the score plots for a selection of test statistics when applied to the modified data set. We observe that for the log-rank  and weighted log-rank test, the shape of scores over time is qualitatively the same as for the original data set, only with lower density at later follow-up times. The same is true for the parametric milestone analysis based on an exponential distribution. For the Kaplan-Meier based tests, however, we can clearly see that early events have become relatively less influential compared to later events. Intuitively this makes sense. For tests that are anchored to a particular (late) timepoint, if there are relatively few events close to that timepoint then they must naturally have a larger influence on the test statistic. The same is true for the flexible parametric milestone analysis. The exponential-model-based milestone analysis avoids this only by making the strong assumption of constant hazard rates. Note that several of the tests have scores that are non-monotonic in time. This demonstrates that non-monotonicity per-se does not imply the test statistic fails to control the type 1 error rate under the strong null hypothesis. The difference here compared to the situation in Figure \ref{fig_fh} is that the non-monoticity is a consequence of censoring. Early events may get lower scores than some later events, but there are now relatively more early events than later events due to censoring. 

\clearpage

\section{Discussion}
\label{sc_discussion}

The choice of appropriate statistical methodology for the primary analysis of a confirmatory clinical trial is always complex. This is especially the case for time-to-event data, and even more so when we anticipate non-proportional hazards. There are many factors to consider. What is the power, and the type 1 error rate, under a range of plausible assumptions? How easy is it to interpret the output? How easy is it to describe the method in a pre-specified analysis plan? Etc. It is unlikely that there is a single method that is uniformly best across all these diverse criteria. Hence the chosen method(s) must be based on a pragmatic compromise.

When looking at the methods described in this paper, there is one clear differentiating factor. The RMST and milestone analyses involve an unambiguous estimand; the weighted log-rank tests do not. Some statisticians argue, as a matter of principle, that the primary analysis of an RCT should be an estimation procedure for an unambiguous single number summary measure \citep{huang2020estimating, snapinn2021comment, paukner2022versatile}. One could argue that the ICH E9 addendum on estimands \citep{ICHE9R1} takes this position for granted, although the main ICH E9 guidance on statistical principles in clinical trials \citep{ICHE9} is more open: 
\textit{"The statistical section of the protocol should specify the hypotheses that are to be tested
and/or the treatment effects which are to be estimated in order to satisfy the primary
objectives of the trial"}.

The argument to focus on estimation rather than testing is strong. Nevertheless, in the specific case of delayed effects in immuno-oncology, the Kaplan-Meier-based RMST and milestone analyses may involve a large loss of power relative to a well-chosen weighted log-rank test. As we have shown above, when early censoring is low, the RMST analysis emphasizes early differences in survival, where treatment differences are expected to be small. A milestone analysis, on the other hand, may lose power owing to an information-leaking dichotomization. A flexible parametric model might solve the latter issue but is difficult to fully pre-specify in a statistical analysis plan. The restriction time $\tau$ is also difficult to pre-specify, owing to uncertainties in the recruitment rate and survival distributions. It may in some circumstances be possible to make the restriction time data dependent \citep{tian2020empirical}, although this complicates the definition of the estimand. Similarly, one might argue that novel estimands such as WMST and AHSW no longer have a simple interpretation.

With the help of the graphical representations introduced in this paper, one can show that certain choices of weighted log-rank test are qualitatively similar to tests based on the estimation of familiar estimands. This helps us to understand what the weighted tests are doing and interpret the results correctly.  The graphical comparisons can be implemented using the \texttt{R} package \texttt{nphRCT} (\cite{nphRCT}).

While helpful, the tools we introduce here do not provide definitive answers on which methods are acceptable for primary analysis in a given context. There are further factors that we have not considered, for example, how compatible are the methods with stratification, covariate adjustment or missing data imputation? Can the methods be implemented in the context of a group-sequential design? Etc. The choice may well come down to which property one is most willing to sacrifice out of robust power, strict type 1 error control, or unambiguous single-number summary statistic.

\section*{Data availability}

Software in the form of an \texttt{R} package, together with a vignette and complete documentation are available on CRAN at \url{https://cran.r-project.org/web/packages/nphRCT/index.html}. Code to reproduce all outputs can be found at \url{https://github.com/dominicmagirr/visualizing_survival_tests.git}.

\bibliographystyle{plainnat}
\bibliography{references}

\end{document}

\maketitle

%%%%%%%%%%%%%%%
\section{From survival data to scaled scores and pseudo-values: step-by-step calculations}

In this section, we illustrate how to compute the scaled scores and scaled pseudo-values using a simple simulated toy-example. In Table \ref{table_toy_data}, we present a toy data set with the time-to-event, arm and whether an event was observed or not. In Table \ref{table_scores_weighted_log_rank} we compute the score of each event time for a weighted log-rank test although, for simplicity, we assume that $w = 1$ which reduces the weighted log-rank test to the standard log-rank test. In Table \ref{table_pseudo_values_rmst} we compute the pseudo-value of each event time for the RMST.

\begin{table}[h]
\centering
\caption{A toy data set consisting of time until the event of interest, treatment group, and presence or not of the event of interest. A lack of presence of the event of interests results in a censored observation.}
\vspace{0.25cm}
\begin{tabular}{|c|c|c|}
\hline
\textbf{Time} & \textbf{Arm} & \textbf{Event} \\ \hline
34.64 & 0 & 0 \\ \hline
4.38 & 0 & 1 \\ \hline
28.69 & 0 & 0 \\ \hline
6.69 & 0 & 1 \\ \hline
24.98 & 0 & 1 \\ \hline
6.32 & 0 & 1 \\ \hline
13.38 & 1 & 1 \\ \hline
33.21 & 1 & 0 \\ \hline
6.12 & 1 & 1 \\ \hline
16.73 & 1 & 1 \\ \hline
27.68 & 1 & 0 \\ \hline
29.46 & 1 & 0 \\ \hline
\end{tabular}
\label{table_toy_data}
\end{table}

As presented in section 3 of the article, a patient with an observed event at time $t_j$ is given a score of $a = w_j - \sum_{i=1}^{j} w_i \times (O_i / n_i)$ whereas if the patient has a censored event, the score $a = \sum_{i=1}^{j} w_i \times (O_i / n_i)$.

Once we have computed column 8 of Table \ref{table_scores_weighted_log_rank}, we define $A = 2 / (\max(a) - \min(a)) = 1.152$ and $B = 1 - A \times \max(a) = -0.056$. The standardized scores are calculated as $a_j \times A + B$, which we can see in column 9 of Table \ref{table_scores_weighted_log_rank}.

Across treatment groups, the average scaled scores are 0.0973 and -0.209. 

\begin{table}[h]
\centering
\caption{Calculated scores for the weighted log-rank test with $w = 1$ (i.e., the standard log-rank test) from the toy data set presented in Table \ref{table_toy_data}.}
\vspace{0.25cm}
\begin{tabular}{|c|c|c|c|c|c|c|c|c|c|c|}
\hline
\textbf{Time ($t_j$)} & \textbf{Arm} & \textbf{\makecell{Patients \\at risk ($n_j$)}} & \textbf{\makecell{Events \\ ($d_j$)}} & \textbf{\makecell{Censored \\ ($l_j$)}} & \textbf{\makecell{Survival \\ prob.}} & \textbf{$w_j$} & \textbf{\makecell{Score \\ ($a_j$)}} & \textbf{\makecell{
Scaled \\ score}} \\ \hline
4.38 & 0 & 12 & 1 & 0 & 1.000 & 1 &  0.917 & 1.000 \\ \hline
6.12 & 1 & 11 & 1 & 0 & 0.917 & 1 &  0.826 & 0.895 \\ \hline
6.32 & 0 & 10 & 1 & 0 & 0.833 & 1 & 0.726 & 0.780 \\ \hline
6.69 & 0 & 9 & 1 & 0 & 0.750 & 1 & 0.615 & 0.652 \\ \hline
13.38 & 1 & 8 & 1 & 0 & 0.667 & 1 & 0.490 & 0.508 \\ \hline
16.73 & 1 & 7 & 1 & 0 & 0.583 & 1 & 0.347 & 0.344 \\ \hline
24.98 & 0 & 6 & 1 & 0 & 0.500 & 1 & -0.820 & 0.152 \\ \hline
27.68 & 1 & 5 & 0 & 1 & 0.417 & 1 & -0.820 & -1.000 \\ \hline
28.69 & 0 & 4 & 0 & 1 & 0.417 & 1 & -0.820 & -1.000 \\ \hline
29.46 & 1 & 3 & 0 & 1 & 0.417 & 1 & -0.820 & -1.000 \\ \hline
33.21 & 1 & 2 & 0 & 1 & 0.417 & 1 & -0.820 & -1.000 \\ \hline
34.64 & 0 & 1 & 0 & 1 & 0.417 & 1 & -0.820 & -1.000 \\ \hline
\end{tabular}
\label{table_scores_weighted_log_rank}
\end{table}

\begin{table}[h]
\centering
\caption{Calculated pseudo-values for the RMST from the toy data set presented in Table \ref{table_toy_data}.}
\vspace{0.25cm}
\begin{tabular}{|c|c|c|c|c|}
\hline
\textbf{Time ($t_j$)} & \textbf{Arm} & \textbf{\makecell{RMST \\ (excluding time $t_j$)}} & \textbf{Pseudo-value ($p$)} & \textbf{\makecell{Scaled \\ pseudo-value}} \\ \hline
34.64 & 0 & 10.678 & 18.00 & -1.000 \\ \hline
4.38 & 0 & 13.402 & 4.38 & 1.000 \\ \hline
28.69 & 0 & 10.678 & 18.00 & -1.000 \\ \hline
6.69 & 0 & 12.940 & 6.69 & 0.661 \\ \hline
24.98 & 0 & 10.678 & 18.00 & -1.000 \\ \hline
6.32 & 0 & 13.014 & 6.32 & 0.715 \\ \hline
13.38 & 1 & 15.370 & 13.38 & -0.322 \\ \hline
33.21 & 1 & 14.446 & 18.00 & -1.000 \\ \hline
6.12 & 1 & 16.822 & 6.12 & 0.745 \\ \hline
16.73 & 1 & 14.700 & 16.73 & -0.814 \\ \hline
27.68 & 1 & 14.446 & 18.00 & -1.000 \\ \hline
29.46 & 1 & 14.446 & 18.00 & -1.000 \\ \hline
\end{tabular}
\label{table_pseudo_values_rmst}
\end{table}

%In order to calculate the pseudo-values link to the test based on the Milestone survival rate, we follow the same process we followed with the pseudo-values link to the test based on the RMST. The only difference is that, instead of calculating the RMST, we calculate the survival probability at, for instance, $\kappa = 18$. In both treatment groups, we have 6 patients and the (non-parametric) survival probability at $\kappa = 18$ is equal to 0.5. These values are obtained using the function \texttt{survfit} from the \texttt{survival} \texttt{R} package. If we remove the event time 34.64 from the data set, which belongs to the control group, the survival probability in the control group is now equal to 0.4. Therefore, following equation \eqref{eq_pseudo_mlst}, the pseudo-value $p$ associated to the observation time 34.64 is equal to $6 \times 0.5 - (6 - 1) \times 0.4 = 1$. Last, once we have all the pseudo-values $p$ associated to the remaining 11 event times from Table \ref{table_toy_data}, we would follow the same standardization process we used in the pseudo-values linked to the test based on the RMST to obtain the standardized pseudo-values.

%%%%%%%%%%%%%%%
% \section{\texttt{nphRCT} R package}

%%%%%%%%%%%%%%% References

\bibliographystyle{plainnat}
\bibliography{references}